\documentclass[twocolumn,secnumarabic,amssymb, nobibnotes, aps, prd]{revtex4-2}

\usepackage{graphicx}%
\usepackage{siunitx}
\usepackage{longtable}%
\usepackage{float}%
\usepackage{xcolor}%
\usepackage{enumerate}%
\usepackage{setspace}%
\usepackage{changepage}   

\begin{document}
	
	\title{Proton probing measurements of filamentary electromagnetic structure in laser ablation of solids}%
	
	\author{J. L. Peebles,$^1$ P. V. Heuer,$^1$ D. H. Barnak,$^1$ Y. V. Zhang,$^1$ J. R. Davies$^1$}%
	\affiliation{$\mathit{^1}$Laboratory for Laser Energetics, University of Rochester, Rochester, New York 14623-1299, USA}
	\email{jpeebles@lle.rochester.edu}
	\date{\today}
	
	\begin{abstract}
		Proton radiography of laser direct-drive spherical implosions has shown anomalous structures that correspond to strong electric or magnetic fields extending throughout the corona. These fields have the ability to affect laser–target interactions and act as an energy sink. To better understand the these fields, simplified experiments were conducted in planar geometry on the OMEGA EP laser at the Laboratory for Laser Energetics. Varying target material, target size, pulse shape, and intensity, and measured the field structure using dual-axis proton radiography and a 4$\omega$ probe. Proton radiographs were analyzed and quantitatively demonstrate that the growth of these features is dominated by laser energy and target Z. The data strongly supports that a secondary instability as a consequence of the expansion driven Weibel instability in these interactions is the primary driver for these fields.
	\end{abstract}
	
	\pacs{numbers!}
	
	\maketitle
\section{Introduction}
Over the previous decades proton radiography has been used to diagnose electric and magnetic fields in various laser plasma interactions. Early experiments by M. Borghesi, C. K. Li and F. Seguin \textit{et al.}\cite{Borghesi2002, Li2009,Rygg2008,Seguin2012,Zylstra2012} used proton radiography to image electric and magnetic field structures around direct-drive spherical implosions and found a variety of features surrounding the laser-irradiated target, including radial structures emanating from the sphere for several millimeters. Years later, other experiments examining the growth of the Rayleigh-Taylor instability in laser-solid interactions found similar features, but in a very different target geometry \cite{GaoThesis,Nilson}.

While these features were discovered, they were not the focus of their respective works and not analyzed in great detail. However, due to their large scale-length and presence in spherical implosions, it is important to ascertain whether these features could have an impact in ICF implosions. Large volume field structures could act as an energy sink, inhibit charged particle transport, and modify the laser-plasma interaction physics. The presence of large electric or magnetic fields has also been shown to confound charged-particle imaging diagnostics \cite{Heuer2025}.

One of the major difficulties presented by this feature is that fact it does not arise in hydrodynamic simulations for the regimes of the given experiments. These features are observed to develop over several nanoseconds, over several millimeters, and appear to be non-neutral structures some tens of microns wide. While this presents a challenge to understanding the mechanism that generates these fields, further dedicated experiments can improve our understanding of the regimes where these effects are important, how they scale, and provide a reference for future, improved simulations.

In this paper we explore some of the previous experimental work performed, disseminate our own experimental findings over a large parameter space and use these experiments to eliminate some possible mechanisms for the formation of filaments, while providing support for others.

\subsection{Features seen in proton radiography for previous experiments}

\begin{figure}[b]
	\includegraphics[width=0.9\columnwidth]{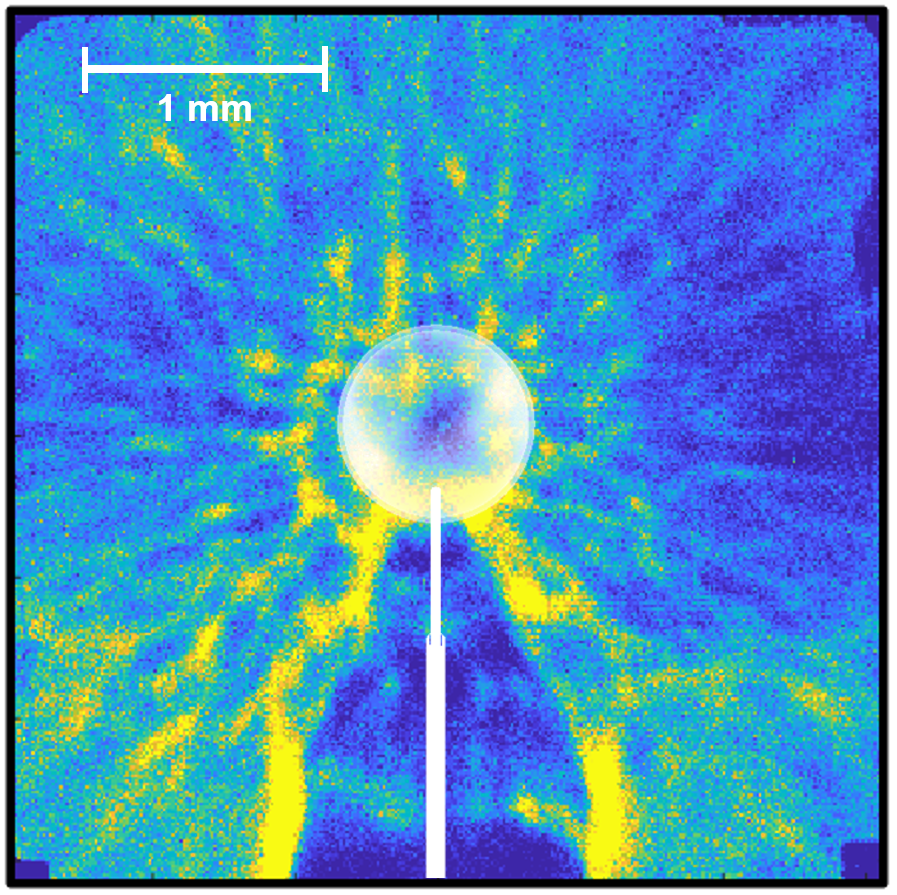}
	\caption{Proton radiograph of a spherical implosion shot on OMEGA demonstrating the generation of web-like electro-magnetic filaments. An drawing of the target (diameter of 860 $\mu m$) prior to being shot is overlayed with the radiograph showing the extent of the filaments in all directions.}
	\label{Fig:OldSpherical}
\end{figure}

Experiments in the spherical regime encompass implosions with shells and shots on solid sphere targets. Proton radiography was used to measure the electromagnetic fields surrounding these implosions. In these experiments there are several distinguishing features; early in time for some shell targets, a ``ripple" like feature developed at a fixed radius from the capsule\cite{Li2009,Igumenshchev}. At later times, filamentary features that look like webbing, extending radially from the targets were observed. These features established themselves within 0.5 ns of the laser turning on and dissipate roughly 1 ns after the drive turns off. The center of the target exhibited focusing, defocusing and scattering effects depending on the experiment. These have not only been correlated to a decrease in proton energy for protons passing through the center of the target, they also appear to greatly affect charged particle diagnostics using high energy deuterons and protons \cite{Heuer2025}. Further experiments conducted later with offset shells show that these features are insensitive to quality of implosion, target offset or even missing many beams (Fig. \ref{Fig:OldSpherical}).

\begin{figure}
	\includegraphics[width=\columnwidth]{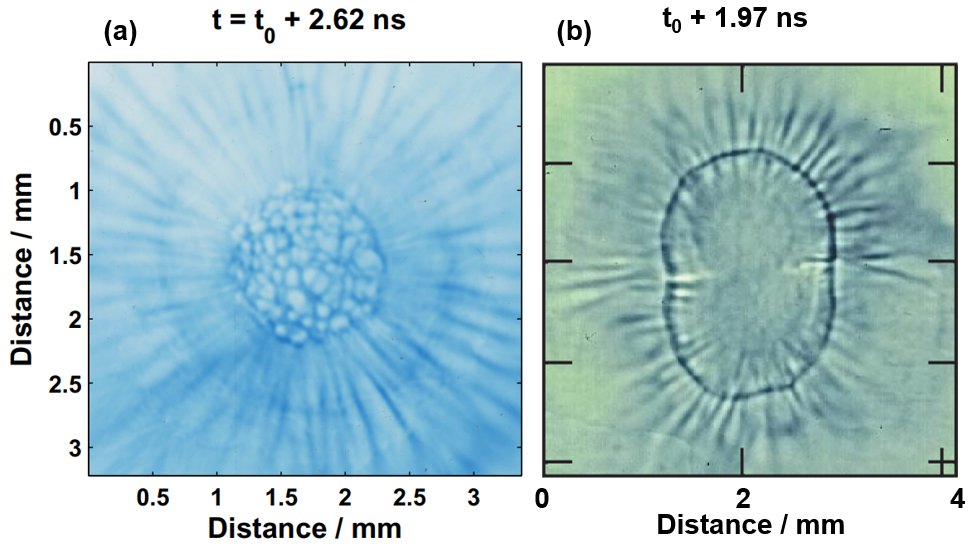}
	\caption{(a) An experiment conducted by L. Gao \textit{et al.}\cite{GaoThesis} examining the electromagnetic fields generated via the Rayleigh--Taylor (RT) instability in thin target interactions. One can see both the lobes and nodes characteristic of RT and the radial filaments that stay coalesced as they move away from the target. (b) A similar experiment conducted by P. Nilson \textit{et al.}\cite{Nilson} examining magnetic reconnection of Biermann Battery ring-like fields between two laser spots. During and after the merge of the Biermann fields, radial filaments can be seen emanating from both spots.}
	\label{Fig:OldPlanar}
\end{figure}

In the planar regime an analogous set of features are shown to evolve on similar time scales\cite{Manuel,Sutcliffe,GaoThesis,Nilson}. However, since the target is cylindrically rather than spherically symmetric, the filaments stay relatively coalesced, rather than breaking up into web like structure. As shown in Gao et al.\cite{GaoThesis} in Figure \ref{Fig:OldPlanar}(a), the laser target interaction zone is surrounded by a strong circular shape indicating a Biermann battery field. The interior structure in the radiograph is characterized by electromagnetic fields from the Rayleigh-Taylor instability, while the outer radial filamentary structure is uncharacterized. Typically in the planar experiment case the filamentary structure was an unwanted feature, which was avoided by either probing early in time, or at lower drive intensity and energy. Reconnection experiments by P. Nilson \textit{et al.}\cite{Nilson} shown in Figure \ref{Fig:OldPlanar}(b) demonstrate similar features. These filamentary features increased the difficulty of measuring the Biermann battery reconnection and further experiments were conducted early in time to avoid the development of these features after 1 ns.

\section{Setup for experiments on OMEGA-EP}

\begin{figure}
	\includegraphics[width=\columnwidth]{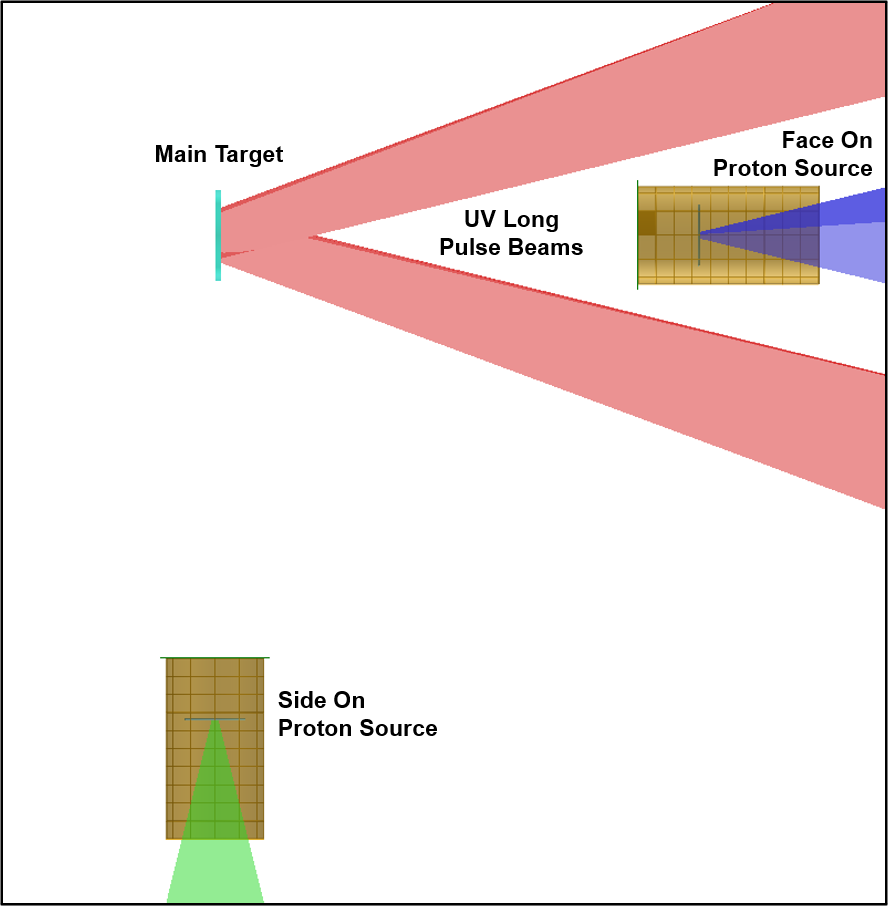}
	\caption{Setup for the majority of experiments: a variable main target is driven by up to two long pulse beams and is probed by a face-on and side-on proton source. The proton sources are driven by the EP sidelighter (green) and backlighter (blue) with up to 300 and 500 J respectively in a 0.7 ps pulse. Protons are deposited on near normal radiochromic film packs 8 cm away from the long-pulse-driven target.}
	\label{Fig:ExpSetup}
\end{figure}

To characterize the filamentary structures extending radially from the target we conducted several experiments using a dual-axis proton radiography platform on the EP laser facility at the Laboratory for Laser Energetics. The experimental setup, shown in Figure \ref{Fig:ExpSetup} utilized up to two long pulse beams at 351 nm, incident on a range of materials (CH, Cu, Si, Mo, Ta and Au), with diameters ranging between 1.0-1.4 mm and thicknesses between 20 and 200 $\mathrm{\mu}$m. The long pulse beams have a variety of distributed phase plates (DPP) available, which smooth the spot and enables the ability to change laser spot size and intensity on target.

These experiments were diagnosed by a dual-axis proton radiography setup using the two orthogonal short pulse beams (1053 nm, 125 - 500 J in a 15 - 20 $\mathrm{\mu}$m spot over 0.7 ps). These beams were incident on 20 $\mathrm{\mu}$m thick copper foils encased in polyether-ether-ketone (PEEK) tubes with a 2 - 5 $\mu m$ Ta shield to protect them from x-rays and other interactions from the long-pulse beams. The protons generated pass by the target and were detected by radiochromic film (RCF) stacks\cite{Roth}. The radiography sources are nominally 5 to 8 mm away from the foil target and probe times account for the roughly 100 -- 200 ps transit time from source to region of interest. The film from the proton radiography was scanned on an EPSON 5000 scanner, typically as 8 bit RGB.

In addition to the proton radiography, a 4$\omega$ probe (263 nm) was utilized to estimate the density structure on the front of the foil target\cite{Froula}. However, the density of the majority of the filamentary structure was too low to be measured by this probe. For several shots a backscatter measurement was taken of the long pulse beams to estimate levels of laser-plasma-interaction effects such as two-plasmon decay (TPD) and stimulated raman scattering (SRS) using the sub-aperture backscatter system (SABS) diagnostic\cite{Filkins}.

\section{Experimental Results}

Typical radiography data is shown in Figures \ref{Fig:Location} and \ref{Fig:EvsB} (a) and (b), which show spoke-like filament caustics (small regions with strong peaks and gradients in proton signal) in the face-on radiographs and two apparent features on the side-on radiographs. In the side-on images, some caustics appear to be in plane with the target, eventually curling off the target surface, while others appear to extend directly outwards along the axis of the target. Features are also seen off the rear surface of the target, but are much weaker and tend to disappear on higher energy radiographs.

The first question raised by the radiographs is whether features in both views can be correlated or whether they are the same features at all. Due to the extent of the filaments, several possible arrangements of fields could generate the caustics. Filaments could be primarily axial with a small angle and small radial extent, only appearing to have a large radial extent due to the increase in magnification as they extend towards the proton source. Filaments can also be potentially radial, leading to the deflections being lost in the target plane of side-on imaging. After the filament geometry can be ascertained, the source of the deflections (magnetic vs electric fields) can be addressed.


\subsection{Field Geometry from Side-On and Face-On Radiography}

Deflections in proton radiography have three typical sources: macroscopic electric or magnetic fields, or scattering by Coulomb collisions. Deflections via electric fields cause focusing or defocusing effects and are independent of probing direction. Magnetic field deflections are very different owing to the $v \times B$ term in the Lorentz force, weakening the dependence on proton probe energy and introducing deflections based on the direction of the probe. Proton scattering in these experiments happens in the dense material of the target, and manifests as a blurring effect.


For deflections due to magnetic fields, probe direction relative to the field lines dictates the type of caustic that is detected. Probing parallel or anti-parallel to a current produces a very strong focusing or defocusing effect, similar to the electric field. However, probing the same field perpendicularly results in a shift in protons, near undetectable without using another fiducial to register proton locations. Most proton radiographs referenced in previous experiments do not use a fiducial and would not be able to detect fields along this direction. Another concern for radiographs of spherically symmetric systems is that any magnetic field strong enough to focus protons will also be strong enough to create a large void if probed from the opposite direction.

\begin{figure}
	\includegraphics[width=\columnwidth]{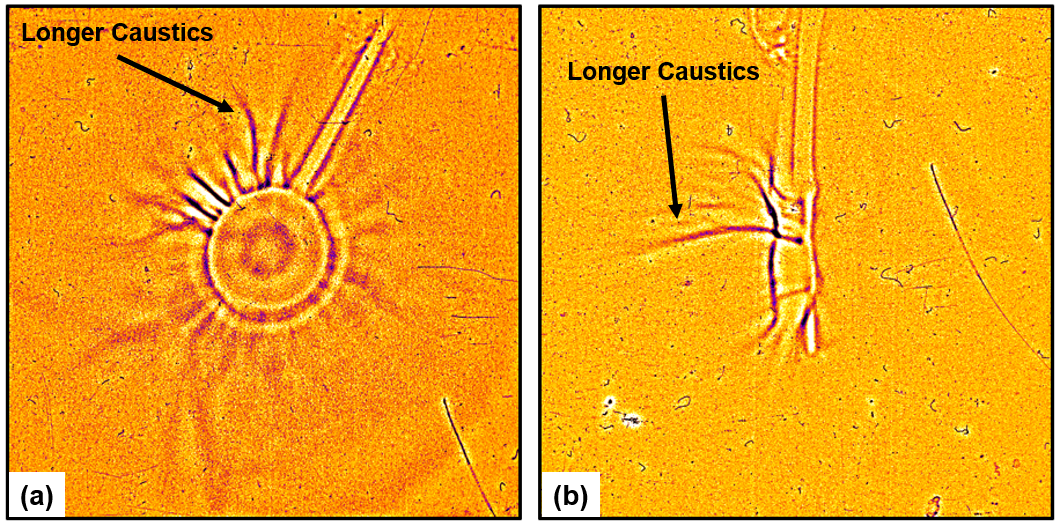}
	\caption{Face-on (a) and side-on (b) radiography view of a Cu target 2.2 ns after start of shot. The relatively strong asymmetry of filaments on this shot allowed for disambiguation of the shape of a single filament. The longer caustic extends approximately 2 mm in both the radial and axial directions.}
	\label{Fig:Location}
\end{figure}

\begin{figure}
	\includegraphics[width=\columnwidth]{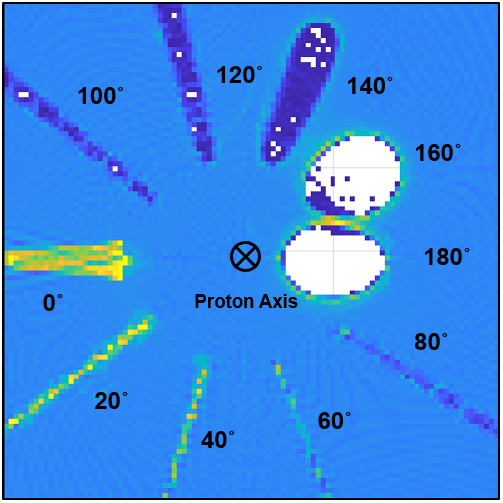}
	\caption{Synthetic proton radiographs of lines of current at different relative angles to the proton axis (into the page). 0 degrees indicates parallel to the proton axis, resulting in a proton focusing effect, while 180 degrees indicates anti-parallel, resulting in a proton void.}
	\label{Fig:Angle}
\end{figure}

This is demonstrated in Figure \ref{Fig:Angle}, where currents of the same magnitude are placed at different angles relative to the probe. One can see that a proton image of an ``axial" current can produce very different results depending on whether the direction is parallel or anti-parallel with the protons. Not only that, a proton image of the same current side-on will see few caustics or deflections. This directional dependency for magnetic fields has the consequence of requiring very specific (or impossible) current geometry, to produce the same type of focusing deflections and caustics in face-on and side-on radiographs simultaneously. Looking at the prior data in Figures \ref{Fig:OldSpherical} and \ref{Fig:OldPlanar}, the filamentary features all appear to be focusing caustics and not voids.

Another aspect to consider is the case of spherical symmetry rather than cylindrical symmetry. Previous experiments in this energy and intensity regime have produced the filamentary structure on spheres\cite{Borghesi2002,Li2009,Seguin2012,Rygg2008}. In spherical geometry current is assumed to be radial owing to the symmetry of the experiment. However, when using radial currents, no geometry is able to reproduce the measured radiographs due to the void effects. An incident proton near the target surface will encounter a current going in the opposite direction at some point, producing a large void structure, especially near the target, which has not been measured in these experiments. This effect was specifically explored in Figure 12 of Heuer 2025 \textit{et. al} \cite{Heuer2025}, where radial currents from a sphere caused substantial voids. This void effect would also occur in planar, cylindrically symmetric experiments in side-on radiographs. While many geometries of magnetic fields and current produce voids in synthetic radiographs, these voids are conspicuously absent in the data, indicating that electrostatic fields are the more likely cause of features seen in experiments.

Further difficulty is encountered when determining the location of the filamentary structure where two lines of sight are not sufficient. However, a few shots developed asymmetric features that could be individually tracked on both radiographs shown in Figure \ref{Fig:Location}. In this case the two views are sufficient and the largest filament can be tracked as a parabolic structure extending 2 mm both radially and axially away from the target in this case. It is reasonable to assume that other similar structures are larger versions of this, extending beyond 4 mm away from the target. On several shots a similar, but inverted parabolic structure can be seen extending outwards from the center of the laser interaction.

Using the parabolic shapes to construct fields from current or distributed charge one can try to reproduce the filaments in radiographs along both lines of sight with either electric or magnetic fields. Since caustics were seen on all films, which can differentiate proton energies up to 60 MeV, field strengths must be sufficient to reproduce caustics for these higher energies. While different proton energy films can be used to differentiate between E and B fields in some cases\cite{Kugland}, this method cannot be applied since the caustics appear on all films.

\begin{figure}
	\includegraphics[width=0.9\columnwidth]{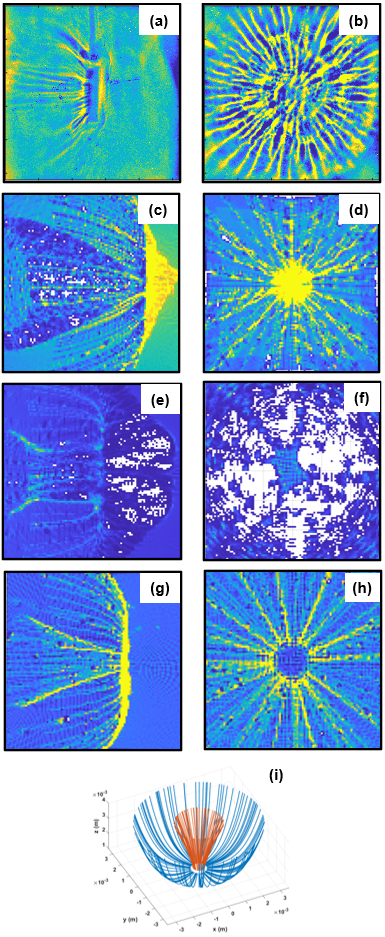}
	\caption{(a) Side-on radiography view of a CH target 2.2 ns after start of shot. (b) Face-on radiography view of the same experiment showing the radially extending filaments. (c,d) Side and face-on synthetic radiographs attempting to reproduce radial filaments with only magnetic fields with concentrated current traveling towards the target and diffuse current traveling away. (e,f) Magnetic field radiographs but with the current directions switched showing significant voids rather than caustics. (g,h) Side and face-on synthetic radiographs attempting to reproduce radial filaments with only electric fields generated by charges placed in parabolic curves. (i) Locations of charge or current for producing synthetic radiographs.}
	\label{Fig:EvsB}
\end{figure}

Best efforts to reproduce the caustics using magnetic fields required 0.5 to 2 kiloAmps of current in parabolic filaments \ref{Fig:EvsB}(i). To achieve the fine focusing of filamentary structure in the face-on radiographs, current going into the target must be concentrated while the current going outwards is diffuse, as a concentrated current coming out of the target would create large voids in the face-on radiographs not seen in experiments (Fig. \ref{Fig:EvsB}(f)). While the field can partially reproduce radial focused filaments as shown in Figure \ref{Fig:EvsB}(d) they over concentrate the protons towards the center of the target. In the side-on radiograph, a large focused region of protons is created close to the target. Since the region closely surrounding the target in side-on radiographs is typically extremely distorted with overlapping caustics(Fig. \ref{Fig:EvsB}(a)), it is hard to say whether any structures seen in synthetic radiographs accurately reproduce the features seen in the side-on case.

The currents used to generate these fields are somewhat contrived since there is no physical explanation as to why current going towards the target should be concentrated. If the current is inverted radiographs in Figure \ref{Fig:EvsB}(e) and (f) are produced which look nothing like the experimental data. Figure \ref{Fig:EvsB}(f) in particular could be checked against an experiment where the drive laser is on the opposite side to the proton probe but requires a non-conventional experimental geometry (facilities rarely place a short pulse opposing a long pulse beam).

Attempting to reproduce the radial filaments using electric fields is a far easier exercise. Spreading tens of nanoCoulombs of negative charge along the parabolic trajectories better reproduces both side-on and face-on radiographs without the convolution required for concentrating currents in a single direction (Fig. \ref{Fig:EvsB}(g,h)). Electric fields are also unaffected by proton probing direction; test shots with a 15 degree angle of the side-on probe relative to the target revealed much of the same filamentary structure, further indicating electric fields are the primary source of deflection. Both the charge and current distributions contained only a fraction of a Joule of energy in the field to create the synthetic radiographs in Figure \ref{Fig:EvsB}.

In summary, while a general comparison between synthetic and experimental radiographs cannot rule out magnetic fields as the cause of the filamentary structure measured, they are most suggestive that they are caused primarily by electric fields.

%

\subsection{Radiograph Analysis, Characterization and Scaling of Filaments}
\subsubsection{Radiograph Analysis}
To better understand the source of the filamentary structure we must be able to quantify aspects of the filaments in the proton radiographs. Proton radiography is difficult to analyze unambiguously owing to the non-uniformity in the proton source, image blurring and ambiguity between electric and magnetic fields. 

\begin{figure}
	\includegraphics[width=\columnwidth]{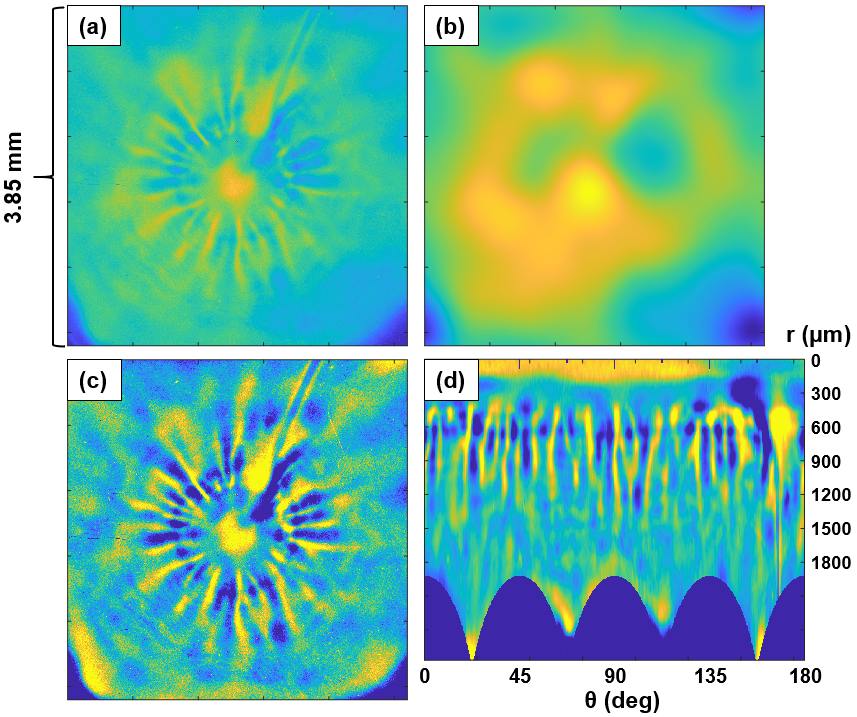}
	\caption{Post processing of radiochromic film scans: (a) A film is scanned on an EPSON 5000 scanner and the red channel of the scan is separated. The image is normalized and background subtracted based on a piece of scanned, undosed film. (b) A low pass butterworth filter is used to estimate low mode perturbations and source profiles. (c) The low modes are divided out of the image revealing the caustics more clearly. (d) The image is transformed into a polar image for further analysis given the cylindrical symmetry of most features.}
	\label{Fig:RCFProcess}
\end{figure}

Figure \ref{Fig:RCFProcess} illustrates the process we apply to analyze the face-on radiographs. The film is scanned and the red color channel is cropped to a 2600x2600 pixel (4600 x 4600 $\mu$m) region for the 8th film in the pack, corresponding to protons of roughly 20 MeV. The image is normalized and background subtracted using a signal free corner of the same film type to account for differences in scanning parameters for some shots. To first overcome the non-uniformity of the proton source we estimate its low mode structure. A low-pass Butterworth filter (second order filter with half power cutoff wavelength of 1.3 mm in the image plane) of the image creates an approximate proton source source before deflection (Figure \ref{Fig:RCFProcess}(b)) which the image is divided by. As shown in Figure \ref{Fig:RCFProcess}(c), removing the non-uniform source helps bring out caustics that were otherwise invisible.

\begin{figure*}
	\includegraphics[width=\textwidth]{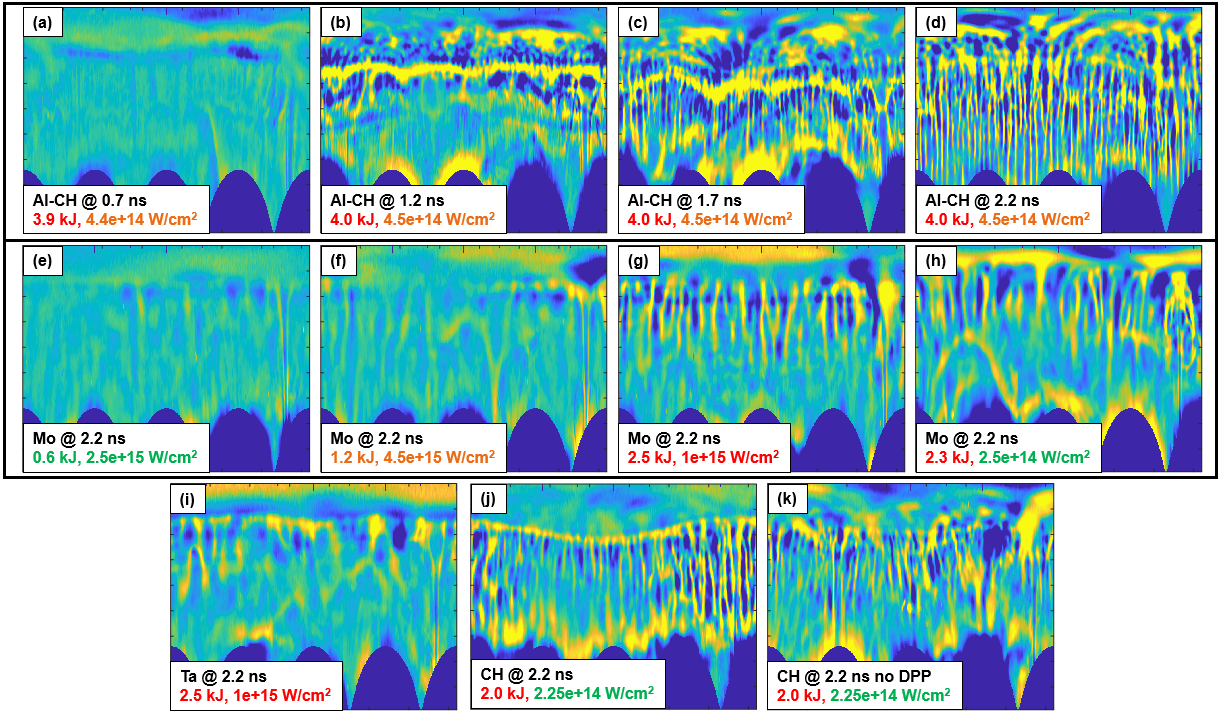}
	\caption{Processed radiographs from a variety of shots. All shots shown here used a 2 ns drive pulse. (a-d): A time series of radiographs taken using four different shots with the same target and laser parameters. Very faint fine scale structure is seen extending for several millimeters, which then grows in strength and lowers in frequency later in time. (e-h): An intensity and energy scan for a mid Z target material (Mo); while it appears that the caustics scale with intensity, an additional shot at low intensity and high energy shows energy to be the determining factor. Higher Z materials also appear to significantly inhibit filament strength. (i) A Tantalum target shot with comparable parameters to (g) showing relative scaling between different target materials. (j,k) Two similar shots on CH with and without a DPP for a single beam drive showing similar caustic strength and growth.}
	\label{Fig:RCFAll}
\end{figure*}

Since the filaments are mostly radial in these radiographs, a polar transformation is applied, shown in Figure \ref{Fig:RCFProcess}(d). Performing a Fourier analysis of bands at the same radius provides the relative strength, angular frequency and radial dependence of the filamentary structure. Total power in the relevant angular modes (60 to 5 deg) can be obtained for each radiograph at several radii and compared across experiments and campaigns.

\subsubsection{Qualitative Scaling of Filaments with Respect to Energy, Intensity and Material}

Since the filaments appear over a wide range of previous experimental campaigns, the key for clarifying the underlying physical cause of the filaments comes from performing multiple shots while changing a single parameter to create a comprehensive database. In conjunction with the radiography analysis one can construct various scalings for how the filaments will grow based on parameters such as material Z, spot smoothness, laser energy and intensity.

Shown in Table \ref{tab:table1} in the appendix is an entire dataset of shots taken on EP with face and side-on radiography with their respective parameters. Shown in Figure \ref{Fig:RCFAll} are selection of face-on radiographs in polar coordinates showing some qualitative results as certain parameters are changed. In the first row is a sequence for the same Al-coated-plastic (100 nm Al on 20 $\mu m$ CH) target shots probed at different times showing the growth rate of the filaments. Fine structure filaments can be seen as early as 0.3 ns, while large structure, stronger strength caustics become more apparent later in time. In the second row are a series of shots with progressively higher intensities or energies on the same, higher Z, material (Mo). The shots on Mo demonstrate that caustic strength scales primarily with energy, especially when comparing the similar radiographs of Figure \ref{Fig:RCFAll} (g) and (h).

A materials scan at similar energies was continued to higher Z, which is shown in Figure \ref{Fig:RCFAll} (g) and (i). These show a decrease in caustic strength and increase in angular wavelength for the filaments as Z increases. Finally spot smoothness was tested by comparing overlapping beams (a-d) to a single beam (j), and to a single beam with the phase plate removed (k), which proved to have a randomizing effect on the filament direction, but similar growth and caustic strength to the other shots. 

\begin{figure}
	\includegraphics[width=0.97\columnwidth]{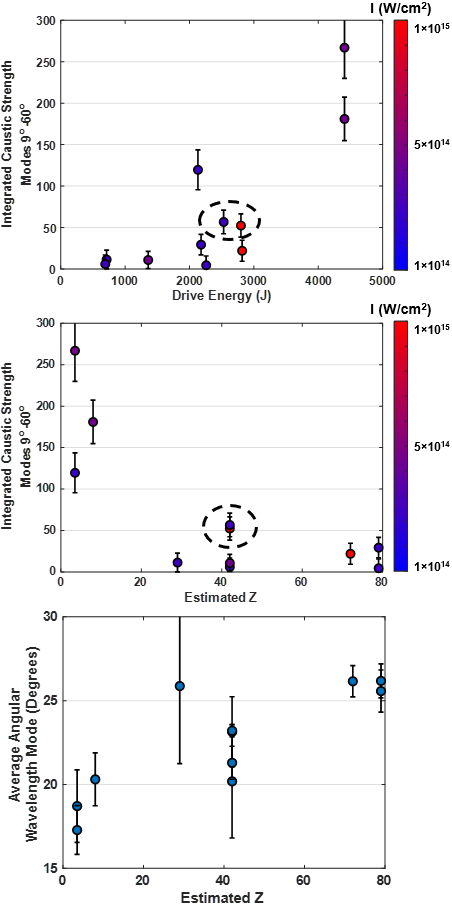}

	\caption{The integrated FFT power spectrum in modes of angular wavelength between 9 and 60 degrees taken at 1.5 mm away from the center of the target with respect to (a) drive energy, (b) estimated material Z. (c) Plots the average angular wavelength mode with respect to increasing Z, showing an increasing wavelength as Z increases. All planar shots from Table I with a probe time of 2.2-2.3 ns and square pulse shape with DPP are included. Of particular interest are the circled shots in (a) and (b) which had a factor of 4 difference in intensity with similar energies, which produced similar filaments}
	\label{Fig:IntegratedModes1}
\end{figure}

To quantify the caustic strength, averaged lineouts of the filtered, polar versions of the radiographs were taken at different radii. Shown in Figure \ref{Fig:IntegratedModes1} is the integrated power spectrum of the lineout taken 1.5 mm from the laser drive for angular wavelengths of interest for shots probed at 2.2 ns. Modes larger than 60 degrees are primarily due to target shape, and modes smaller than 9 degrees can be dominated by small target defects or the stalk. Errors were calculated by taking lineouts over a wider band of radii and using the spread in their integrated power spectra. 

This data shows how caustic strength scales with respect to drive energy, material Z and drive intensity. There is clearly a general trend of increasing strength as drive energy increases and material Z decreases. Intensity appears to have minimal impact, especially when comparing the two shots circled with similar energy and identical material, but different intensity. Average angular mode wavelength was shown to be increase with material Z, which can also be seen qualitatively when comparing the polar images from Figure \ref{Fig:RCFAll}(d), (g) and (i).

In summary the data collected demonstrate that the filament growth rate is highly dependent on laser energy and material Z, while being less dependent on laser intensity and independent of laser smoothness. The extent of the filaments decreases with material Z for the same conditions, while the angular wavelength increases.

\section{Discussion of Possible Mechanisms}

As described in section III the proton radiography from each of these experiments is highly indicative of electric fields as the primary deflection force, which is reconcilable in both radiography fields of view. However, a strong electric field alone is irreconcilable with the spatial and time scales that these features can exist. After all, if such electric fields existed on their own, the plasma would quickly move to neutralize the filamentary features, since in the area far from the target, collision time and plasma frequencies are still far faster than the time scale that the features are observed. The time scale of these features suggest that a magnetic field must play a role in governing and maintaining the filamentary structure, even if they are not the primary source of proton deflection.

One potential source of magnetic field is the magneto-thermal instability (MTI) \cite{Tidman}, where plasma temperature perturbations perpendicular to density gradients can lead to Righi-Leduc heat flow and generation of fields azimuthally around the gradients. However, the circumstances for these gradients to arise are not consistent with our experiments. The MTI requires a large system size that is inconsistent with steady state solution with fixed density and temperature gradients. Simulations analyzed by F. Garcia-Rubio \textit{et al}\cite{Fernando} found that the parameter space where the MTI could arise was extremely limited, and would certainly not occur as ubiquitously as has been seen in our or other experiments.

\begin{figure}
	\includegraphics[width=\columnwidth]{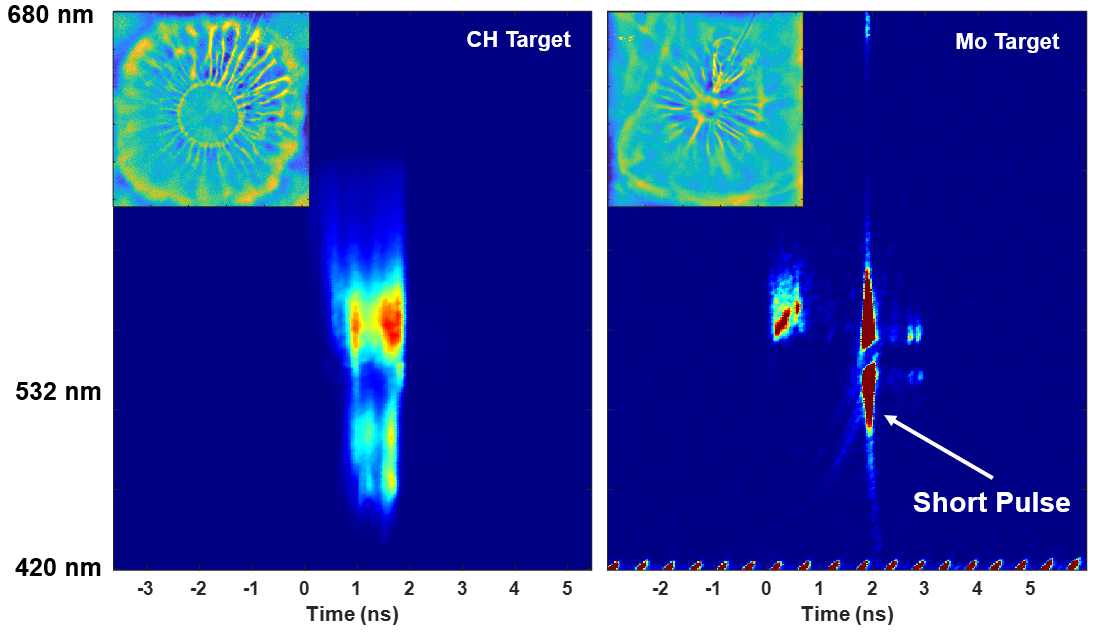}
	\caption{Sub-aperture Backscatter (SABS) measurements taken of scattered light from the drive beam on two shots. (Left) A shot on a CH target with an intensity of 2.5 $\times 10^{14}$ W/cm$^2$ showing significant scattered light as a consequence of laser-plasma interactions characteristic of SRS and generated many filaments. (Right) A similar shot on a Mo target at the same intensity which generated many filaments but minimal scattered light right at the beginning of the pulse, indicating SRS is not a primary driver of the anisotropy.}
	\label{Fig:SABS}	
\end{figure}

Another commonly observed mechanism for generating the magnetic fields needed for filaments to develop is the Weibel instability \cite{Weibel}.The Weibel instability requires an anisotropic velocity distribution of particles in order to grow, which can come from a multitude of potential sources. One aspect initially explored in this study was the possibility of a hot electron population accelerated by various laser plasma interaction mechanisms, either Stimulated Raman Scattering (SRS)\cite{Masson,Silva,Zhao,Zhou} or Two Plasmon Decay (TPD). Since the filamentary structure seemed to evolve close to the SRS scalings with respect to intensity and material Z, measurements were taken of the backscattered light as shown in Figure \ref{Fig:SABS}. While these initially supported the notion that LPI was causing an anisotropic distribution of electrons, a specific shot Figure (\ref{Fig:RCFAll} (h)) specifically lowered intensity below the SRS limit while maintaining high energy on target. This showed similar filamentary growth to a comparable high intensity shot in Figure \ref{Fig:RCFAll} (g), indicating that LPI actually had little role in driving the electron anisotropy required for the development of the filaments. This conclusion is also supported by experiments by F. Seguin \textit{et. al}\cite{Seguin2012}, who dropped intensity for spherical implosions to the point where no SRS or TPD was observable, but still measured filaments in their radiographs.

A second potential source of anisotropy seeding the Weibel instability are sharp temperature gradients in the plasma. Temperature gradients that are much larger than the electron collisionless skin depth, but smaller than the plasma mean free path (non-local or gradients would be damped quickly) can generate magnetic fields perpendicular to the temperature gradient\cite{Schoeffler1, Schoeffler2}. In the experimental regimes where the filaments were measured this anisotropy would most likely manifest in a plasma of several keV, where gradients would be a few microns in size. The only source of such gradients would be the speckle and axial filamentation of the laser through the plasma as it reaches near critical density. While this could generate localized Weibel B fields on the order of the laser spot size, this cannot account for the extent of the fields far outside the laser path where density is far lower and there are no such temperature gradients. One would also expect that our experiments without a phase plate or overlapped beams would see significant changes in field growth, but this did not appear to be the case. However, while these fields may start in a localized region, it is possible that they are pushed outwards from the target later in time.

Finally, a third potential source of anisotropy comes from the expansion of the plasma causing a difference in temperature along the expansion direction compared to its perpendicular counterparts. This expansion driven Weibel has been explored by K. V. Lezhnin, who has done the first simulations of this effect with laser heating and collisions\cite{LezhninArxiv,LezhninDPP}, following from previous work by C. Thaury, V. V. Kocharovsky, \textit{et al,} \cite{Thaury,Kocharovsky}. The expansion driven Weibel effect initially appears to not align with the measured instability; since the expansion is in the direction away from the surface, the anisotropy that would develop is best described as $T_\perp > T_\parallel$, leading to magnetic fields growing perpendicular to a wavevector parallel to the expansion direction (radially). This would manifest as rings or ripples outwards from the target. If the anisotropy is inverted ($T_\perp < T_\parallel$) magnetic field features growing in spokes radially can be created, however, there is no known mechanism to lead to this situation arising spontaneously in the plasma.

However, it is possible that the fields from the expansion driven Weibel are the seed for a secondary effect causing the radial filaments. Such an effect was hypothesized decades ago \cite{Kalman,Stockem,Taggart}, and has been found in simulations in the context of the Weibel instability \cite{DelSarto,Du}. These articles found that only focusing elements were typically observed in proton radiographs, rather than the void structure that could originate with magnetic field alone. Simulations showed that while both fields were present, the deflections from the electric field were less energy intensive compared to those from magnetic field. Work by Dieckmann \textit{et al.}\cite{Dieckmann} demonstrated that after the Weibel instability grows, the electric field of coalesced charge and magnetic pressure become balanced, described by $E_\parallel = -\nabla|B_\perp|^2/e\mu_0n_e$.

\begin{figure}
	\includegraphics[width=\columnwidth]{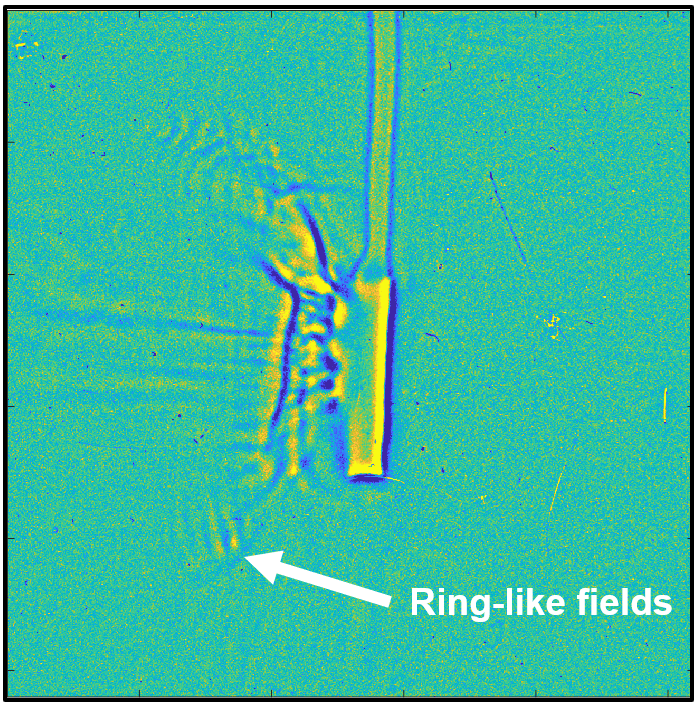}
	\caption{Radiograph taken from a gold target shot at 2.2 ns. Slight ripple or ring-like features expand outwards in a cone from the target, consistent with the geometry of expansion driven Weibel instability. These rings are not present in the face-on radiography for this shot.}
	\label{Fig:RCFRings}
\end{figure}

Furthermore, faint ring-like fields were occasionally witnessed in our experiments and others as shown in Figures \ref{Fig:OldSpherical}, \ref{Fig:RCFProcess}(c) and \ref{Fig:RCFRings}. Such features have also been seen in proton radiography of spherical implosions such as in Igumenshchev (2014) \cite{Igumenshchev}, manifesting as thin ripple-like features around the target. In cylindrical implosions, ripple-like features were measured around the target traveling radially outwards\cite{HeuerPRad}. In the planar experiments these fields cause weaker caustics more readily visible in the side-on radiography. The fact that the observation of these features is geometry dependent indicates that they are likely cased by magnetic fields rather than electric. The wave vector of these fields are perpendicular to the expansion direction consistent with the expansion driven Weibel theory.

In this way the expansion driven Weibel instability generates a seed for a secondary electric field that arises from electrons passing through it. This electric field is perpendicular to the magnetic field, which places it back in line with the expansion direction of the plasma and is consistent with our data.

We can further investigate whether the Weibel instability is expansion driven by comparing the growth rates of expansion driven Weibel simulations to our data. In this way we can ascertain whether factors such as laser energy, intensity or target material have an expected effect. In Lezhnin \textit{et al.} the growth rate of the collisionless Weibel instability based on plasma parameters is:
\begin{equation}
	\gamma_{nocoll} = \frac{2}{3\sqrt(3\pi)}\frac{v_{th,e}\omega_{pe}}{c}\frac{A^{3/2}}{A+1}
\end{equation}

where anisotropy is described by $A \equiv T_{e\perp}/T_{e\parallel} - 1$ where perpendicular and parallel are with respect to the expansion direction, and $v_{th,e}$ and $\omega_{pe}$ are the electron thermal velocity and plasma frequency.

\begin{figure}
	\includegraphics[width=0.90\columnwidth]{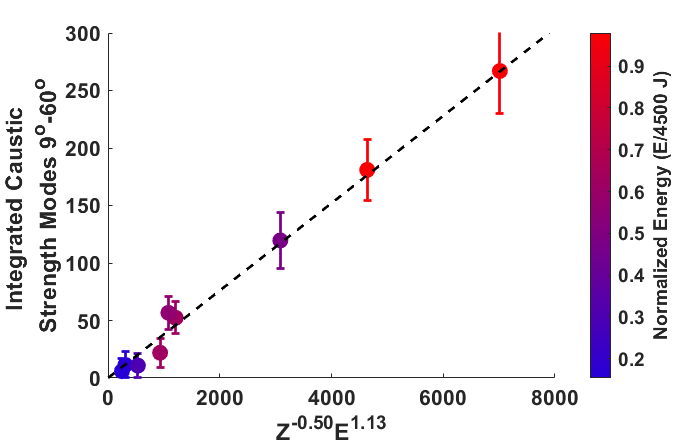}
	\caption{Integrated FFT data from figure \ref{Fig:IntegratedModes1} plotted as a scaling of $E^{1.13}/\sqrt{Z}$.}
	\label{Fig:IntegratedModes2}
\end{figure}

This growth rate was then used to create a threshold parameter based on experimental parameters such as laser intensity and target material using a steady-state laser expansion model by W. M. Manheimer \textit{et al.}\cite{Manheimer}, which considered laser intensity and energy the same since these were 1D simulations.
\begin{equation}
	\Gamma = 2\frac{\mu^{2/3}}{Z^{5/3}}\bigg(\frac{\lambda}{1\ \mathrm{\mu m}}\bigg)^{11/3}\bigg(\frac{\mathrm{I}}{10^{13}}\bigg)^{4/3}
\end{equation}

where $\mu$ and $Z$ are the material atomic mass and number, $\lambda$ is laser wavelength normalized to a micron wavelength and I is laser intensity normalized to $10^{13}$.

In reality only a portion of the growth rate is indirectly dependent on laser intensity through the scaling of electron temperature, while plasma expansion in 2D or 3D is based primarily on laser energy. However, importantly, collisionality and plasma frequency of the system are based on the plasma density, which is primarily driven by the energy in the laser and mass of the material, not intensity at these scales.

These findings are consistent with the change in filament strength found in our radiographs as we scaled energy, material Z, intensity and distance. As shown in Figure \ref{Fig:IntegratedModes1} it was found that the growth of these structures was highly dependent on material Z and incident beam energy. Attempting to fit both of these parameters to a scaling it was found that in the region just outside the target caustic strength scaled with $\sqrt{Z}$ and $E^{0.5-1.1}$ with less dependency on energy further from the target. Very weak dependence on intensity (with constant energy) was found due to the low-intensity, high-energy shot on Mo, which produced the same filamentary strength as a similar shot with a smaller spot size and same energy. In planar geometry difficulty arises in finding consistent scalings at multiple distances from the center of the target due to the hemispherical nature of plasma expansion and because density depends on all three parameters: material Z, incident energy and distance from interaction region. However, the fact that the feature scaling is strongly coupled to plasma expansion further supports expansion driven Weibel as a seeding instability for the features measured.

\section{Summary and Conclusions}
Through careful experimentation and exploration of the parameter space we have discovered several important aspects related to the filamentary caustics measured with proton radiographs of spherical implosions. First, the features measured in most experiments are likely to be a secondary electrostatic instability seeded by the presence of a magnetic field generated by the expansion-driven Weibel instability. While a magnetic field may be able to produce some features measured in the experiment, they require very specific probing directions and asymmetries in the plasma.

As demonstrated, the deflections measured are much more likely to be caused by an electric field sustained by the magnetic field, which is more energetically feasible and more robust to probing direction. The fields inferred from the radiography are negligible in terms of energy storage indicating that this field structure cannot account for an substantial energy sink effect in ICF implosions. This, however, does not mean that such fields can not impact laser energy coupling through various LPI mechanisms beyond the scope of this work.

Second, the instabilities are highly dependent on material Z and incident laser energy and only weakly dependent on laser intensity. Unfortunately, not only does this means that the instability is difficult to mitigate by dropping the intensity below a threshold on target, the instability will substantially increase with experiment size and driver energy. With regards to most direct-drive ICF experiments, which will typically use most of the energy available to them, these filaments are unavoidable. As shots with Tantalum and Molybdenum targets demonstrated, even moving to higher Z may not entirely remove filaments, especially as drive laser energy increases.

\section*{Appendix}
\begin{table*}[!htbp]
	\caption{\label{tab:table1}
		Shot parameters for entire dataset. Shots without a Distributed Phase Plate (DPP) were focused to 750 um diameter.}
	\begin{ruledtabular}
		\begin{tabular}{c c c c c c c c c}
			\textrm{\# of Beams}&
			\textrm{Pulse Length)}&
			\textrm{Energy}&
			\textrm{Intensity}&
			\textrm{DPP}&
			\textrm{Target Mat}&
			\textrm{Target Size}&
			\textrm{Probe Time}&\\
			\textrm{}&
			\textrm{(ns)}&
			\textrm{(J)}&
			\textrm{$10^{14}$(W/cm$^2$)}&
			\textrm{}&
			\textrm{(Thickness $\mu m$)}&
			\textrm{}&
			\textrm{(ns)}&\\
			\colrule
			& & & & & & & &\\
			2 & 2 & 4007 & 4.54 & SG8-750 & Al(0.1)-CH(20) & 1.4 mm square & 2.3 &\\
			2 & 2 & 4042 & 4.58 & SG8-750 & Al(0.1)-CH(20) & 1.4 mm square & 1.7 &\\
			2 & 2 & 4007 & 4.54 & SG8-750 & CH(20) & 1.4 mm disk & 1.2 &\\
			2 & 2 & 3959 & 4.48 & SG8-750 & Al(0.1)-CH(20) & 1.4 mm square & 3.4 &\\
			2 & 1.8 shaped & 3702 & 4.66 & SG8-750 & Al(0.1)-CH(20) & 1.4 mm square & 2.3 &\\
			2 & 1.8 shaped & 3653 & 4.60 & SG8-750 & Al(0.1)-CH(20) & 1.4 mm square & 1.5 &\\
			2 & 2 & 3916 & 4.43 & SG8-750 & CH(20) & 1.4 mm disk & 0.6 &\\
			1 & 2 & 1937 & 2.19 & SG8-750 & CH(100) & 1.4 mm disk & 2.3 &\\
			1 & 2 & 1991 & 2.25 & None & CH(100) & 1.4 mm disk & 2.3 &\\
			1 & 2 & 2053 & 2.32 & SG8-750 & Au(3)-CH(100) & 1.4 mm disk & 2.2 &\\
			1 & 2 & 2014 & 2.28 & SG8-750 & CH(20) & 0.86 Sphere & 2.2 &\\
			1 & 2 & 1990 & 2.25 & SG8-750 & CH(100) & 1.4 mm disk & 0.5 &\\
			1 & 2 & 1981 & 2.24 & SG8-750 & Au(80) & 1.4 mm disk & 2.2 &\\
			
			2 & 2 & 647 & 2.58 & SG10-400 & Cu(25) & 1.4 mm disk & 2.3 &\\
			2 & 2 & 629 & 2.50 & SG10-400 & Mo(25) & 1.4 mm disk & 2.2 &\\
			2 & 2 & 2544 & 1.01 & SG10-400 & Mo(25) & 1.4 mm disk & 2.2 &\\
			2 & 2 & 2561 & 1.02 & SG10-400 & Ta(25) & 1.4 mm disk & 2.2 &\\
			2 & 2 & 1234 & 4.91 & SG10-400 & Mo(25) & 1.4 mm disk & 2.2 &\\
			
			2 & 2 & 4602 & 5.21 & SG8-750 & Mo(25) & 1.0 mm disk & 0.6 &\\
			2 & 2 & 4534 & 5.13 & SG8-750 & Mo(25) & 1.0 mm disk & 0.3 &\\
			1 & 2 & 2300 & 2.60 & SG8-750 & Mo(25) & 1.0 mm disk & 2.3 &\\
			2 & 2 & 4467 & 5.06 & SG8-750 & Mo(25) & 1.0 mm disk & 1.0 &\\
			1 & 2 & 2296 & 2.60 & None & Mo(25) & 1.0 mm disk & 2.3 &\\
			1 & 2 & 2273 & 2.57 & None & Mo(25) & 1.0 mm disk & 0.5 &\\
			1 & 2 & 2134 & 2.42 & SG8-750 & CH(25) & 1.0 mm disk & 0.3 &
		\end{tabular}
	\end{ruledtabular}
\end{table*}
	
	\section*{Acknowledgments}
	This material is based upon work supported by the Department of Energy [National Nuclear Security Administration] University of Rochester “National Inertial Confinement Fusion Program” under Award Number(s) DE-NA0004144

	This report was prepared as an account of the work sponsored by an agency of the U.S. Government. Neither the U.S. Government nor any agency thereof, nor any of their employees, makes any warranty, expressed or implied, or assumes any legal liability or responsibility for the accuracy, completeness, or usefulness of any information, apparatus, product, or process disclosed or represents that its use would not infringe privately owned rights. Reference herein to any specific commercial product, process, or service by trade name, trademark, manufacturer, or otherwise does not necessarily constitute or imply its endorsement, recommendation, or favoring by the U.S. Government or any agency thereof. The views and opinions of the authors expressed herein do not necessarily state or reflect those of the U.S. Government or any agency thereof.

\end{document}